# Implementation of atomically defined Field Ion Microscopy tips in Scanning Probe Microscopy

**William Paul, Yoichi Miyahara, and Peter Grütter**

Department of Physics, Faculty of Science, McGill University, Montreal, Canada.

E-mail: paulw@physics.mcgill.ca

**Abstract**
The Field Ion Microscope (FIM) can be used to characterize the atomic configuration of the apex of sharp tips. These tips are well suited for Scanning Probe Microscopy (SPM) since they predetermine SPM resolution and electronic structure for spectroscopy. A protocol is proposed to preserve the atomic structure of the tip apex from etching due to gas impurities during the transfer period from FIM to SPM, and estimations are made regarding the time limitations of such an experiment due to contamination by ultra-high vacuum (UHV) rest gases. While avoiding any current setpoint overshoot to preserve the tip integrity, we present results from approaches of atomically defined tungsten tips to the tunneling regime with Au(111), HOPG, and Si(111) surfaces at room temperature. We conclude from these experiments that adatom mobility and physisorbed gas on the sample surface limit the choice of surfaces for which the tip integrity is preserved in tunneling experiments at room temperature. The atomic structure of FIM tip apices is unchanged only after tunneling to the highly reactive Si(111) surface.

## 1. Introduction

The utility of the Field Ion Microscope (FIM) for characterizing tips destined for Scanning Tunneling Microscopy (STM) experiments was first considered by Fink [1] soon after the invention of the STM. The main advantage of an atomically defined tip apex is clear: If the exact atomic arrangement of the apex is known precisely, the electronic structure of the tip and the lateral resolution of the STM are predetermined. The same detailed knowledge of the apex termination, as well as the tip radius, is useful for Atomic Force Microscopy (AFM) experiments, and important in the interpretation of results in combined STM and AFM experiments [2,3].

Furthermore, since the transport properties of a molecular junction are sensitive to the atomic arrangement of the contact electrodes [4], FIM characterized tips are excellent candidates for constructing and studying well-defined junctions in STM/AFM.

We have previously used FIM characterized tips in combined STM/AFM nanometer-scale indentation experiments. Since their radius of curvature is known with high precision, it is possible to gain insights regarding nanoscale plasticity and junction conductivity which could not otherwise be established [5,6].

Atom Probe Field Ion Microscopy (AP-FIM), a variant of FIM which measures the mass-to-charge ratio of field evaporated tip material, has been used to investigate material transfer in tunneling, voltage pulsing, and indentation experiments[7–11]. These, and other combined FIM/STM studies [12–15] shed light on some of the atom transfer processes in the tip-sample junction. A detailed study presenting a protocol for the preservation of tip apex structures in SPM experiments still does not exist.

In this work we focus on the challenges involved in preserving the atomic structure of the very apex of the prepared tip against corrosion by undesirable but inevitable rest gas molecules in ultra-high vacuum (UHV) and by impurities in the helium imaging gas. Then, we discuss the approach of these tips to a tunneling interaction distance with the surface of Au(111), highly ordered pyrolytic graphite (HOPG), and Si(111) samples. These are difficult experimental challenges, but can be overcome by the proposed protocol and good vacuum practice.

The procedures described herein for implementing FIM tips in SPM experiments are additionally valuable considering the recent progress in producing ultra-sharp tips by techniques such as gas etching and faceting by noble metal plating [16–21]. The popularity of etched tungsten tips for AFM experiments based on qPlus sensors [22] and length-extension resonators [23] opens up new possibilities for implementing atomically defined FIM tips in SPM experiments [24,25].

## 2. Tip apex preparation

Tungsten tips are fabricated from polycrystalline and single-crystalline (111) wire by DC electrochemical etching [26–28]. Polycrystalline tips nearly always terminate with a (110) oriented grain at the apex (to within a few degrees) due to the crystallographic texture arising from the cold drawing process of bcc metals [29,30]. The tips are annealed in UHV by resistive heating and field emission cycles.

Before SPM experiments, the tips are cleaned in FIM by removing a few atomic layers by field evaporation. This is achieved by increasing the DC imaging voltage by about 10% relative to the FIM imaging voltage at room temperature. By monitoring the apex structure of a W(111) tip during field evaporation, a trimer (3-atom) tip can be prepared by lowering the applied voltage when this structure is achieved. The trimer tip is a common stable apex termination for W(111) tips with radii in the range of 3-12 nm.

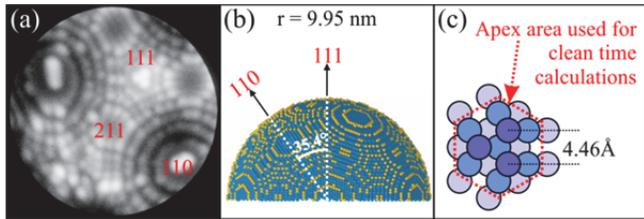

**Figure 1:** (a) typical W(111) trimer tip prepared by field evaporation, imaged at 5.8kV; (b) ball model (side view) of a 9.95nm radius tip, illustrating (111) and (110) directions; (c) trimer apex model illustrating the area used in calculations of clean time.

A typical W(111) trimer tip is shown in Figure 1(a). The radius is determined to be 10.3±0.7 nm by the ring counting method [27,31] (see appendix for details). A ball model is shown in Figure 1(b) with a radius of 9.95nm (this specific radius was chosen to reproduce the trimer apex in the model).

In this study, we use W(110) tips to characterize the contamination of tip apices by imaging gas, while W(111) tips are used for all STM tunneling studies. The tips are all characterized and maintained at room temperature throughout the FIM and STM experiments. They are stable and observable for hours at room temperature.

## 3. Tip apex preservation

*3.1. Imaging gas impurities*

Dedicated FIM systems intended for diffusion studies have very stringent vacuum requirements. They are often made of glass to minimize the presence of hydrogen and are designed for high temperature bakeout (limited to 300°C by the micro-channel plate (MCP)) [32]. These systems are usually baked out several times, and the MCP is bombarded for many days by 200eV electrons to remove gas trapped in the channels. Gases are admitted through Vycor or quartz tubes and titanium getters and cold traps are used extensively. These systems must be clean enough to avoid rest gas atoms landing on the atomic planes being studied for diffusion during the experiment. Due to the complexity of the combined FIM-SPM apparatus, the system cannot be as thoroughly clean as the dedicated FIM systems described.

In our FIM studies, helium is used as an imaging gas. It is admitted through a heated quartz tube to provide an ultra-clean source of helium. However, due to the slow leak rate of the quartz tube, the system cannot be pumped actively with a turbomolecular pump. All pumping during FIM is done passively with a liquid nitrogen cooled titanium sublimation pump (TSP). Diffusing the helium gas through heated quartz introduces minimal impurities into the vacuum system, but the imaging gas purity is still limited by the outgassing of the vacuum system itself.

Regardless of the source of the imaging gas (quartz tube, or leaked in directly from a gas line), impurities are likely to be in higher concentration while the gas is admitted. It is important to keep these impurities from reacting with the tip apex atoms during FIM and while pumping the imaging gas after FIM. We demonstrate the very rapid etching of the tip apex due to imaging gas impurities by preparing a clean tungsten tip in Figure 2(a) and momentarily lowering and raising the imaging voltage from 5.1kV to 0V and back. This voltage ramping process took 40s.

The many changes to the atomic structure of the tip are highlighted in Figure 2(b) in a manner employed by Müller to identify individual changes among the many atomic sites of a tip [33]. A colour superposition image is constructed by illuminating the initial image in green and the final image in red. They are aligned and re-exposed (digitally in Matlab) to produce the superposition image. Areas that show up in green correspond to atoms that have field evaporated, whereas atoms that show up in red correspond to adatoms that have appeared. Missing atom sites (green) should also be considered as sites where an adsorbed gas atom had previously been bound: Many adsorbed species weaken the bonding of underlying tungsten atoms and remove them when they field desorb (which may happen before an adequate field is reached to produce a helium FIM image) [34,35].

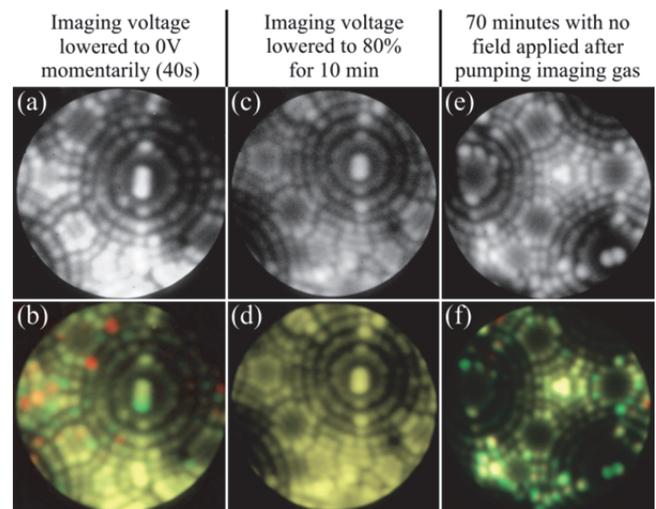

**Figure 2:** (a) (110) tip at 5.1 kV; (b) colour superposition image of (a) with the FIM image after the imaging voltage was momentarily lowered and raised – about 30 atomic sites show modifications; (c) (110) tip at 5.0 kV; (d) colour superposition image of (c) with the FIM image after waiting 10 minutes with the voltage lowered to 4.0 kV – no changes are observed to the atomic structure; (e) (111) tip at 6.0kV; (f) colour superposition image of (e) with the FIM image after waiting 70 minutes in UHV.

The ~30 changes visible on the tip's ~120 nm² surface area in ~40 seconds are discouraging to the prospect of preserving the atomically defined tip apices. Fortunately, helium has a much higher ionization energy than all other gases [36], and as long as a sufficient field is applied to the tip, other gas species will be ionized with a greater rate and accelerated away from the tip before they can react with the atoms on the apex. Experimentally, we find that reducing the tip voltage to about 80% of the imaging voltage is sufficient to keep contaminant gases from reacting with the tip during extended periods of time. The somewhat reduced magnitude of the field helps ensure that no atomic changes to the tip occur due to field evaporation. We call this the "force field" and demonstrate it on a tip in Figure 2(c) which is exposed to the imaging gas and its impurities at 80% of the imaging voltage. Ten minutes later, no changes are observable in the colour superposition micrograph of Figure 2(d).

The "force field" method should be applied to ensure that the tip apex remains clean between FIM and SPM. The "force field" voltage is applied to the tip while the helium is pumped from the chamber and maintained until the UHV system returns to base pressure. Processes that result in pressure transients such as flashing the TSP and refilling its liquid nitrogen can be done while the "force field" is on to avoid affecting the tip apex.

After performing a scanning probe measurement with the FIM tip, it is necessary to perform FIM again to check the integrity of the apex. During the He filling process, we are faced with the choice of whether or not to apply the "force field". If the "force field" is applied before turning on the MCP and leaking in helium, field desorption events occurring at voltages lower than the "force field" voltage will not be imaged, and cannot be counted. Working quickly with a fresh TSP shot, we have found that it is possible to keep the tip reasonably clean during the FIM start-up sequence without the "force field". This way, the number of adsorbed atoms during the experiment interval may be overestimated, but field desorption events below the "force field" voltage can be recorded.

To summarize, we propose the following FIM protocol to preserve the tip apex from imaging gas impurities in SPM experiments:
(1) Prepare a clean tip by field evaporation in FIM
(2) Reduce applied voltage to ~80% ("force field")
(3) Pump imaging gas, refill liquid nitrogen cold traps, flash TSP while "force field" is on
(4) Ramp down "force field" voltage when pressure has recovered
(5) SPM experiment

## 3.2. UHV rest gas

During the interval of the SPM experiment, the FIM tip will be subject to contamination by the UHV rest gas. The vacuum requirements necessary to keep the apex of a tip clean during such an experiment are now considered. The trimer apex of a (111) tip is shown schematically in Figure 1(c). A dashed hexagon is drawn around the perimeter the first and second layer of a (111) trimer tip. We consider the likelihood of rest gas molecules impinging on this 1.16nm² area. In this discussion, we address only the probability of various gas species impinging on the apex. Many subtle effects regarding sticking coefficients are at play such as the anisotropic adsorption coverage of gases on a tip surface due to the difference of dissociation rates on different atomic planes [37]. The following estimations represent an upper limit of tip contamination as they do not include adsorption probabilities.

The rest gas composition in UHV is first estimated using a residual gas analyzer (RGA) in our preparation chamber. The location of the RGA is different from the location of the FIM/SPM measurements, but we assume that the gas composition will be similar for this similarly pumped vacuum chamber (with the possible overestimation of H₂O partial pressure because of the more thorough FIM/STM chamber bakeout). The gas composition is calculated by the SRS RGA Windows software, and is presented in Table 1. For a total pressure reading $P_{gauge} = 5 \times 10^{-11}$ mbar, the individual partial pressures $p_i$ of the component gases are calculated as

$$p_i = n_i \frac{P_{gauge}}{\sum_i \frac{K_i}{K_{N_2}} n_i} \qquad (1)$$

where $n_i$ is the pressure fraction of the gases and $K_i/K_{N_2}$ is the ionization sensitivity [38] of the gauge to the various gases.

The flux of each gas species can be calculated from their respective masses and partial pressures by combining the Boltzmann velocity distribution of the gas particles with the ideal gas law [39]:

$$F_i = \frac{p_i}{\sqrt{2\pi m_i kT}} \qquad (2)$$

Given a flux per unit time and area, $F$, and assuming that the arrival of gas molecules is independent of time and location, the probability of obtaining $k$ arrivals in an area $A$ in time $\tau$ is given by a Poisson distribution [40]:

$$P(k) = \frac{e^{-FA\tau}(FA\tau)^k}{k!} \qquad (3)$$

We are interested in the probability of obtaining more than one arrival, i.e. $k > 1$, which is easily obtained using the normalized nature of the distribution:

$$P(k > 1) = \sum_{k=1}^{\infty} P_k = 1 - P_0 = 1 - e^{-FA\tau} \qquad (4)$$

We describe the rate of contamination of the tip apex area in 2 ways: In the first, we calculate the duration of an experiment such that 5% of experiments of this duration would count more than one gas molecule impinging on the apex. The second is the probability of more than one gas molecule impinging on the apex during a 60 min delay. The results of these calculations are shown in Table 1. Note that other unidentified gases (not listed) make up 2.2% of the gas composition, meaning they have approximately the same statistics as $N_2$.

| Species | Mass, $m$ (amu) | Ionization Sensitivity, $K_i/K_{N_2}$ | Pressure Fraction, $n_i$ | Partial Pressure, $p_i$ (mbar) | Flux, $F$ (nm$^{-2}$ s$^{-1}$) | 1:20 apex contamination time(min) | P(k>1) gas atom on apex during a 60 min delay |
|---|---|---|---|---|---|---|---|
| H2 | 2 | 0.42 | 88.0% | 9.4E-11 | 1.0E-03 | 0.7 | 98.5% |
| H20 | 18 | 0.9 | 4.6% | 4.9E-12 | 1.8E-05 | 41 | 7.1% |
| N2 | 28 | 1 | 2.3% | 2.4E-12 | 7.0E-06 | 100 | 2.9% |
| CO | 28 | 1.2 | 1.7% | 1.8E-12 | 5.2E-06 | 140 | 2.1% |
| CO2 | 44 | 1.4 | 1.1% | 1.2E-12 | 2.7E-06 | 270 | 1.1% |
| He | 4 | 0.16 | 0.1% | 1.1E-13 | 8.1E-07 | 890 | 0.3% |

Table 1: Table of typical UHV rest gas species, their partial pressures, fluxes, and calculated contamination times and probabilities for a 1.16 nm² tip apex.

Due to its low mass and large pressure fraction, hydrogen has a high flux and correspondingly very large probability of contaminating tips. In addition, hydrogen gas cannot be imaged in FIM and is known to have no corrosive behavior on tungsten tips [35], so its presence goes undetected with the methods employed here.

The next most prevalent gas species ($H_2O$, $N_2$, and CO) are at a nearly acceptable background level if an experiment is completed within a short period of time (~1 hour). These species are also known to etch tungsten atoms, so changes to the atomic structure would likely be detectable in FIM [34,35,37,41]. However a more convincing confirmation of a contamination-free tip would be the experimental repeatability of the electronic properties of its tunneling junction with an atomically clean surface.

We illustrate typical rest gas contamination of a tip in Figure 2 (e) which was preserved with a "force field" until the UHV system had returned to base pressure. The "force field" was then turned off, and the tip was left in UHV for 70 minutes (a reasonable SPM experiment duration). The colour superposition image in Figure 2 (f) shows ~30 changes to atomic sites. The number of changes is overestimated due to the contamination during the FIM start-up sequence. Assigning an approximate mass of 28 amu, the changes to the tip during this interval indicate and effective contaminant pressure of ~$2 \times 10^{-11}$ mbar. This number is consistent with the calculations in Table 1 and the expected overestimation of contamination due to the FIM start-up sequence.

From these rest gas calculations, it is clear that experiments must be performed quickly, even in a clean UHV environment. This has important consequences for the design of a combined FIM/SPM experiment: Since the sample should not be subjected to contamination from the imaging gas, the tip or sample must be transferred in a reasonable time frame, and the coarse approach of the tip to the sample surface must be done hastily.

## 4. Tunneling gap formation

Having established a protocol for preserving FIM tip apices, and showing that they can be kept atomically clean for a reasonable length of time, we now consider the challenges in forming the tunneling gap between a FIM tip and a surface. The approach of FIM tips to Au(111), HOPG and Si(111)-2x1 surfaces at room temperature is presented.

The initial coarse approach is monitored optically, and the tip can be brought to 5-10 μm from the surface by observing its reflection off the sample.

The bandwidth of our tunneling current preamplifier is ~2 kHz [42], and the response time of the GXSM controller [43] feedback loop is also set on the order of ~1 ms. Since tunneling currents typically increase at a rate of about one decade per Angstrom, it is necessary to ensure that the approach speed does not exceed 1000 Å/s so that the feedback loop can detect and correct the tip position without serious overshoot. We manage to achieve negligible current overshoot with a tip approach speed of ~200 Å/s.

We prepare trimer tip apices before each experiment by field evaporation (with the exception of a rather asymmetric apex used in one experiment which supported unusually stable tetramer tips). The tip is brought to tunneling interaction with the sample within 30-50 minutes of the time the "force field" voltage is turned off. Once in tunneling range, we wait about one minute for z-piezo creep to settle, and then perform small 10x10 nm scans to check the sample slope. The tip is maintained within tunneling distance of the sample for 4-5 minutes before it is withdrawn and FIM is performed again (images are acquired with lower applied voltage to capture adsorbed atoms). The tunneling current throughtout the entire experiment is recorded at a sample rate of 50 kHz (with an external 8[th] order antialiasing filter at 10 kHz) using a National Instruments USB-6259 BNC controlled by Matlab.

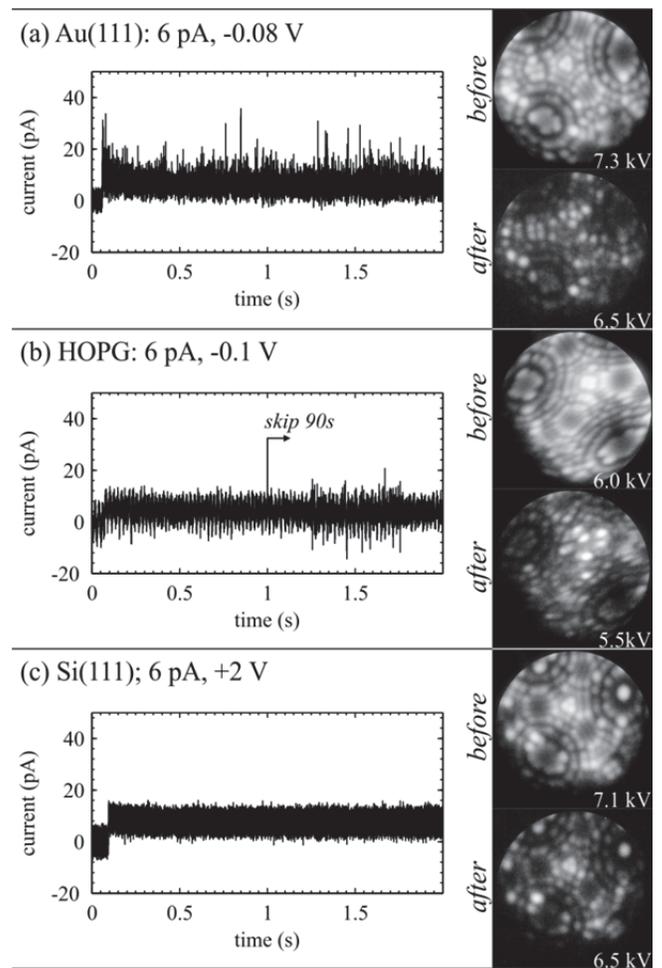

**Figure 3:** FIM tip apices before and after, as well as current traces upon approach to (a) Au(111), (b) HOPG, and (c) Si(111) surfaces.

The approach of an atomically defined tip to a Au(111) surface is illustrated in Figure 3(a). The Au(111) surface consists of a 100 nm Au(111) film epitaxially grown on mica. It is prepared in UHV by Ne[+] ion sputtering and annealing cycles, to several cycles beyond the disappearance of all carbon contamination detected by Auger electron spectroscopy. We routinely prepare Au(111) and observe a well-ordered herringbone reconstructed surface in STM. As soon as the tunneling gap is established (6 pA, -0.08 V on the sample), many spikes are seen on the tunneling current up to ~40pA. The spikes are continuous throughout the experiment, and this behavior is representative of the 4 other experiments of approaching atomically defined W(111) tips to Au(111). After the FIM tip is withdrawn from the tunnel junction, the apex is severely modified by adsorbed atoms.

HOPG was chosen as an alternative substrate to determine the cause of the tip apex contamination. The samples were cleaved in air and hastily inserted into vacuum and then degassed overnight at 150ºC in UHV to remove physisorbed gases. The tunneling current (6 pA, -0.1 V on the sample) behaviour on HOPG was generally much quieter, but during the two experiments on HOPG, occasional bursts of noise were observed, such as the one ~90 s into the experiment, shown in Figure 3(b). Again, the FIM tips withdrawn from the tunnel junction were covered with adsorbates.

Clean Si(111)-2x1 surfaces were prepared by dicing a groove half-way through a Si(100) wafer, mounting it at a 35.3º angle such that the (111) cleavage plane was normal to the sample holder, and cleaving in UHV less than 30 minutes before the tunneling junction was established. The cleaved surfaces have terraces of variable width depending on the measurement position relative to the tensile or compressive edge of the cleave, ranging from 2-50 nm. A trimer tip was prepared in FIM and approached until a current of 6 pA at +2 V sample bias was achieved. This experiment was performed with two tips on two different cleaved Si(111) samples, and no spikes in tunneling current were detected, which is apparent in the current trace in Figure 3(c). The tip apex returned unchanged from the tunneling junction.

## 5. Discussion

We suspect that the return of contaminated tip apices from Au(111) results from mobile adatoms on the sample surface at room temperature. The presence of a STM tip can lower the barrier for diffusion of adatoms toward the tip and for atom transfer to the tip [44–46]. Another possible source of the adsorbed atoms that had to be considered is physisorbed gas diffusing from the tip shank (collected during FIM cycles) to the tip-sample junction along the field gradient. Fortunately, it seems that the source of the tip contamination is not from the tip shank since the tips approached to Si(111) return with identical apices.

HOPG was investigated because of its stronger covalent bonding within the plane, and therefore very different surface diffusion properties. Unfortunately the preparation of HOPG by cleavage in air followed by degassing in UHV was probably not sufficient to form an atomically clean surface free of diffusing contaminants, which ultimately changed the FIM tip apex structure. Unfortunately the exact species of the transferred atoms cannot be determined in our system.

The result of unaltered FIM tips after tunneling to Si(111) is not surprising. The surface is very reactive and should therefore be free of diffusing adatoms. Its preparation by cleavage in UHV immediately before the experiment also guarantees that it is free of contamination.

The implementation of atomically defined tips in SPM experiments at room temperature is therefore also limited by the choice of substrate. Spikes in the tunneling current signal correlate with changes in tip apex structure in FIM. We also conclude that changes to the atomic structure of the tip are due to adatoms or physisorbed gas molecules transferred from the sample, and not from the tip shank.

## 6. Conclusion

A protocol is introduced for preserving the tip apex of FIM tips using a "force field". This serves to keep the apex unchanged while pumping the FIM imaging gases, and carrying out dirtier vacuum processes. The time limitations for performing an experiment in typical UHV rest gas conditions without apex contamination were estimated to be reasonable for all gas species except molecular hydrogen. Hydrogen is not observable in FIM and desorbs without changing the underlying tip structure.

Using the protocol established here, atomically defined FIM tips were prepared and carefully approached to tunneling proximity with Au(111), HOPG, and Si(111) surfaces. It was shown that the atomic structure of FIM tips is destroyed by material transferred from the sample on Au(111) and HOPG. At room temperature, Si(111) could be approached without modifying the atomic structure of the tip apex, indicating that the material observed in FIM after tunneling experiments on Au(111) and HOPG originated from the sample. This also suggests a careful choice of substrate is necessary when attempting experiments with atomically defined tips at room temperature. The drastically changed tip apices returning from tunneling experiments with Au(111) and HOPG point to the fact that STM at room temperature in UHV is a very dynamic process involving atom transfer even at very small currents.

Guidelines for implementing FIM tips in SPM experiments have been established, as well as the first convincing demonstrations of tip-sample interactions without modification of the atomic geometry of tip apices as determined in FIM.

## Acknowledgments

Funding from NSERC, CIFAR, and RQMP is gratefully acknowledged.

## Appendix

It is important to point out that the ring counting method is commonly applied inaccurately in the estimation of FIM tip radii. We now discuss where this inaccuracy arises: To proceed with ring counting, we choose a crystallographic direction $(hkl)$ and count the number of rings $n$ on the FIM image between $(hkl)$ and $(h'k'l')$. Assuming a spherical envelope of the tip shape, and having calculated the plane spacing for planes perpendicular to $(hkl)$ to be $s_{(hkl)}$, the local radius of curvature is given by $R = ns_{(hkl)}/(1 - \cos\theta)$ where $\theta$ is the angle between directions $(hkl)$ and $(h'k'l')$.

Starting at the (111) pole in Figure 1(a) and working over to the (110) pole, we count about 8 rings. A common error is to insert $n = 8$ into the above equation, along with the plane spacing $s_{(111)} = 0.912$ Å for the (111) direction. As illustrated by the arrows in Figure A.1, near the (111) pole, there exists one (111) plane per ring, but as the angle increases toward (110), there are many more (111) planes per counted ring. The edges of individual (111) planes are no longer visible in FIM because they make up the smooth surface of the close packed (110) plane. This leads to a significant underestimation of $n$, given our choice of $s_{(111)}$.

Carefully examining the ball model in Figure A.1, it is apparent that there is a one-to-one correspondence between the rings and (110) planes. An accurate estimation of the tip radius is achieved by using $s_{(110)} = 2.23$ Å and counting rings from the (110) axis to the (111) or (211). *Estimations from the ring counting method are only valid when the rings being counted correspond to single atomic steps*.

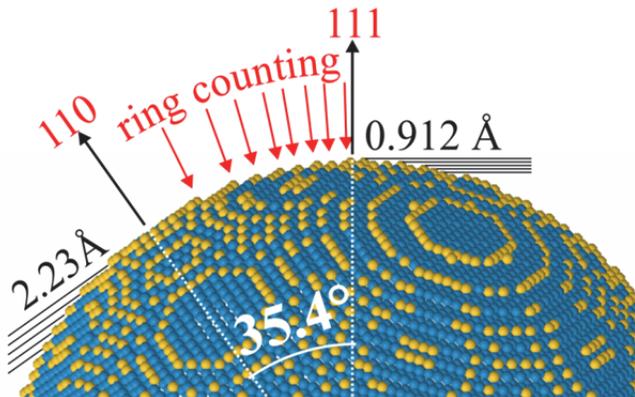

**Figure A.1**: Ball model of a 9.95nm radius tip, illustrating (110) and (111) poles and interatomic plane spacings. Rings counted in the FIM image between these poles are indicated with arrows.

The average radius determined from the 6 pairs of (110)-(111) and (110)-(211) planes visible in the micrograph shown in Figure 1(a) is 10.3±0.7 nm. The quoted uncertainty represents the standard deviation of the obtained radii. The features of this tip correspond well with those of the 9.95 nm ball model reconstruction.

To quickly estimate a tungsten tip radius in nm, one can simply multiply the number of rings from (110) to (211) by 1.66, or multiply the number of rings from (110) to (111) by 1.21 (these factors are $s_{(110)}/(1 - cos\theta)$). This inspires the mnemonic: "One one oh to one one one? Multiply by one two one!"

## References

[1] Fink H-W 1986 Mono-atomic tips for scanning tunneling microscopy *IBM J Res Dev* **30** 460–165
[2] Welker J and Giessibl F J 2012 Revealing the Angular Symmetry of Chemical Bonds by Atomic Force Microscopy *Science* **336** 444–9
[3] Ternes M, González C, Lutz C, Hapala P, Giessibl F J, Jelínek P and Heinrich A J 2011 Interplay of Conductance, Force, and Structural Change in Metallic Point Contacts *Physical Review Letters* **106** 1–4
[4] Mehrez H, Wlasenko A, Larade B, Taylor J, Grütter P and Guo H 2002 I-V characteristics and differential conductance fluctuations of Au nanowires *Physical Review B* **65** 1–13
[5] Cross G L W, Schirmeisen A, Grutter P and Dürig U 2006 Plasticity, healing and shakedown in sharp-asperity nanoindentation *Nat Mater* **5** 370–6
[6] Oliver D J, Maassen J, El Ouali M, Paul W, Hagedorn T, Miyahara Y, Qi Y, Guo H and Grütter P 2012 Conductivity of an atomically-defined interface *(in preparation)*
[7] Weierstall U and Spence J C H 1998 Atomic species identification in STM using an imaging atom-probe technique *Surface Science* **398** 267–79
[8] Spence J C H, Weierstall U and Lo W 1996 Atomic species identification in scanning tunneling microscopy by time-of-flight spectroscopy *Journal of Vacuum Science & Technology B: Microelectronics and Nanometer Structures* **14** 1587
[9] Fian A and Leisch M 2003 Study on tip–substrate interactions by STM and APFIM *Ultramicroscopy* **95** 189–97
[10] Fian A, Ernst C and Leisch M 1999 Combined atom probe and STM study of tip-substrate interactions *Fresenius' Journal of Analytical Chemistry* **365** 38–42
[11] Wetzel A, Socoliuc A, Meyer E, Bennewitz R, Gnecco E and Gerber C 2005 A versatile instrument for in situ combination of scanning probe microscopy and time-of-flight mass spectrometry *Review of Scientific Instruments* **76** 103701
[12] Kuk Y and Silverman P J 1986 Role of tip structure in scanning tunneling microscopy *Applied Physics Letters* **48** 1597
[13] Sakurai T 1990 New versatile room-temperature field ion scanning tunneling microscopy *Journal of Vacuum Science & Technology A: Vacuum, Surfaces, and Films* **8** 324
[14] Tomitori M, Hirano N, Iwawaki F, Watanabe Y, Takayanagi T and Nishikawa O 1990 Elaboration and evaluation of tip manipulation of scanning tunneling microscopy *Journal of Vacuum Science & Technology A: Vacuum, Surfaces, and Films* **8** 425
[15] Nishikawa O, Tomitori M and Iwawaki F 1992 High resolution tunneling microscopies: from FEM to STS *Surface Science* **266** 204–13
[16] Kuo H-S, Hwang I-S, Fu T-Y, Hwang Y-S, Lu Y-H, Lin C-Y, Tsong T T and Hou J-L 2009 A single-atom sharp iridium tip as an emitter of gas field ion sources. *Nanotechnology* **20** 335701
[17] Chang C-C, Kuo H-S, Tsong T T and Hwang I-S 2009 A fully coherent electron beam from a noble-metal covered W(111) single-atom emitter. *Nanotechnology* **20** 115401
[18] Rezeq M, Pitters J and Wolkow R 2006 Tungsten nanotip fabrication by spatially controlled field-assisted reaction with nitrogen *The Journal of Chemical Physics* **124**
[19] Pitters J L, Urban R and Wolkow R a 2012 Creation and recovery of a W(111) single atom gas field ion source. *The Journal of chemical physics* **136** 154704
[20] Rahman F, Onoda J, Imaizumi K and Mizuno S 2008 Field-assisted oxygen etching for sharp field-emission tip *Surface Science* **602** 2128–34
[21] Bryl R and Szczepkowicz A 2006 The influence of the oxygen exposure on the thermal faceting of W[111] tip *Applied Surface Science* **252** 8526–32
[22] Giessibl F J 1998 High-speed force sensor for force microscopy and profilometry utilizing a quartz tuning fork *Applied Physics Letters* **73** 3956–8
[23] Heike S and Hashizume T 2003 Atomic resolution noncontact atomic force/scanning tunneling microscopy using a 1 MHz quartz resonator *Applied Physics Letters* **83** 3620
[24] Song Y J, Otte A F, Shvarts V, Zhao Z, Kuk Y, Blankenship S R, Band A, Hess F M and Stroscio J 2010 Invited Review Article: A 10 mK scanning probe microscopy facility. *The Review of scientific instruments* **81** 121101
[25] An T, Eguchi T, Akiyama K and Hasegawa Y 2005 Atomically-resolved imaging by frequency-modulation atomic force microscopy using a quartz length-extension resonator *Applied Physics Letters* **87** 133114
[26] Hagedorn T, El Ouali M, Paul W, Oliver D, Miyahara Y and Grütter P 2011 Refined tip preparation by electrochemical etching and ultrahigh vacuum treatment to obtain atomically sharp tips for scanning tunneling microscope and atomic force microscope. *The Review of scientific instruments* **82** 113903
[27] Lucier A-S, Mortensen H, Sun Y and Grütter P 2005 Determination of the atomic structure of scanning probe microscopy tungsten tips by field ion microscopy *Physical Review B* **72** 1–9
[28] Lucier A-S 2004 *Preparation and Characterization of Tungsten Tips Suitable for Molecular Electronics Studies, M.Sc. Thesis* (McGill)
[29] Verlinden B, Driver J, Samajdar I and Doherty R D 2007 *Thermo-Mechanical Processing of Metallic Materials* (Oxford: Elsevier)
[30] Greiner M and Kruse P 2007 Recrystallization of tungsten wire for fabrication of sharp and stable nanoprobe and field-emitter tips *Review of Scientific Instruments* **78** 026104
[31] Tsong T T 1990 *Atom-probe field ion microscopy* (New York: Cambridge University Press)
[32] Antczak G and Erlich G 2010 *Surface Diffusion* (New York: Cambridge University Press)
[33] Müller E W 1957 Study of Atomic Structure of Metal Surfaces in the Field Ion Microscope *Journal of Applied Physics* **28** 1

[34] Ehrlich G and Hudda F G 1962 Direct Observation of Individual Adatoms: Nitrogen on Tungsten *The Journal of Chemical Physics* **36** 3233

[35] Mulson J F and Müller E W 1963 Corrosion of Tungsten and Iridium by Field Desorption of Nitrogen and Carbon Monoxide *The Journal of Chemical Physics* **38** 2615

[36] Lias S G 2011 Ionization Energies of Gas-Phase Molecules *CRC Handbook of Chemistry and Physics*

[37] Ehrlich G and Hudda F G 1960 Observation of Adsorption on an Atomic Scale *The Journal of Chemical Physics* **33** 1253

[38] Singleton J H 2001 Practical guide to the use of Bayard–Alpert ionization gauges *Journal of Vacuum Science & Technology A: Vacuum, Surfaces, and Films* **19** 1712–9

[39] Weston G F 1985 *Ultrahigh Vacuum Practice* (Toronto: Butterworth & Co.)

[40] Zelterman D 2006 *Models for Discrete Data* (New York: Oxford University Press)

[41] Ehrlich G and Hudda F G 1963 Promoted field desorption and the visibility of adsorbed atoms in the ion microscope *Philosophical Magazine* **8** 1587–91

[42] Dürig U, Novotny L, Michel B and Stalder A 1997 Logarithmic current-to-voltage converter for local probe microscopy *Review of Scientific Instruments* **68** 3814

[43] Zahl P, Wagner T, Möller R and Klust A 2010 Open source scanning probe microscopy control software package GXSM *Journal of Vacuum Science & Technology B: Microelectronics and Nanometer Structures* **28** C4E39

[44] Sørensen M, Jacobsen K and Jónsson H 1996 Thermal Diffusion Processes in Metal-Tip-Surface Interactions: Contact Formation and Adatom Mobility *Physical Review Letters* **77** 5067–70

[45] Li J, Berndt R and Schneider W-D 1996 Tip-Assisted Diffusion on Ag(110) in Scanning Tunneling Microscopy *Physical Review Letters* **76** 1888–91

[46] Kürpick U and Rahman T 1999 Tip Induced Motion of Adatoms on Metal Surfaces *Physical Review Letters* **83** 2765–8